\documentclass[sigconf]{acmart}

\makeatletter
\let\authorOrig\author
\renewcommand{\author}[1]{\ignorespaces}
\let\affiliationOrig\affiliation
\renewcommand{\affiliation}[1]{\ignorespaces}
\let\authornoteOrig\authornote
\renewcommand{\authornote}[1]{\ignorespaces}

\renewcommand{\authornotemark}[1][]{\ignorespaces}
\let\orcidOrig\orcid
\renewcommand{\orcid}[1]{\ignorespaces}
\let\cityOrig\city
\renewcommand{\city}[1]{\global\@ACM@citypresenttrue\ignorespaces}
\let\countryOrig\country
\renewcommand{\country}[1]{\global\@ACM@countrypresenttrue\ignorespaces}
\renewcommand{\email}[2][]{\ignorespaces}
\let\institutionOrig\institution
\renewcommand{\institution}[1]{\global\@ACM@instpresenttrue\ignorespaces}
\makeatother

\usepackage{xcolor}
\usepackage{cleveref}

\definecolor{ggred}{HTML}{F8766D}
\definecolor{ggblue}{HTML}{619CFF}
\definecolor{gggreen}{HTML}{00BA38}
\definecolor{ggpurple}{HTML}{B983FF}
\definecolor{ggorange}{HTML}{FF9E4A}
\definecolor{ggpink}{HTML}{FB61D7}
\definecolor{ggyellow}{HTML}{FFC61E}
\definecolor{ggteal}{HTML}{00BFC4}
\colorlet{ggredlight}{ggred!60}
\colorlet{ggbluelight}{ggblue!60}
\colorlet{gggreenlight}{gggreen!60}
\colorlet{ggpurplelight}{ggpurple!60}
\colorlet{ggorangelight}{ggorange!60}
\colorlet{ggpinklight}{ggpink!60}
\colorlet{ggyellowlight}{ggyellow!60}
\colorlet{ggteallight}{ggteal!60}

\long\def\rev#1{{\color{black}#1}}

\makeatletter
\newsavebox{\highlightbox}
\newdimen\highlightwd
\newdimen\highlightht
\newdimen\highlightdp
\newdimen\highlighttotalht
\newdimen\highlightradius
\newdimen\highlightcurve
\newcommand{\highlight}[2][yellow!40]{%
  \begingroup
  \sbox{\highlightbox}{#2}%
  \highlightwd=\wd\highlightbox
  \advance\highlightwd by 5pt%
  \highlightht=\ht\highlightbox
  \advance\highlightht by .9pt%
  \highlightdp=\dp\highlightbox
  \advance\highlightdp by .9pt%
  \highlighttotalht=\highlightht
  \advance\highlighttotalht by \highlightdp
  \highlightradius=1.2pt%
  \highlightcurve=.663pt%
  \leavevmode
  \ifdefined\pdfliteral
    \raisebox{-\highlightdp}{%
      \hbox to \highlightwd{%
        \rlap{{\color{#1}%
         \pdfliteral{q
           \strip@pt\highlightradius\space 0 m
           \strip@pt\dimexpr\highlightwd-\highlightradius\relax\space 0 l
           \strip@pt\dimexpr\highlightwd-\highlightradius+\highlightcurve\relax\space 0\space
           \strip@pt\highlightwd\space \strip@pt\dimexpr\highlightradius-\highlightcurve\relax\space
           \strip@pt\highlightwd\space \strip@pt\highlightradius\space c
           \strip@pt\highlightwd\space \strip@pt\dimexpr\highlighttotalht-\highlightradius\relax\space l
           \strip@pt\highlightwd\space \strip@pt\dimexpr\highlighttotalht-\highlightradius+\highlightcurve\relax\space
           \strip@pt\dimexpr\highlightwd-\highlightradius+\highlightcurve\relax\space \strip@pt\highlighttotalht\space
           \strip@pt\dimexpr\highlightwd-\highlightradius\relax\space \strip@pt\highlighttotalht\space c
           \strip@pt\highlightradius\space \strip@pt\highlighttotalht\space l
           \strip@pt\dimexpr\highlightradius-\highlightcurve\relax\space \strip@pt\highlighttotalht\space
           0\space \strip@pt\dimexpr\highlighttotalht-\highlightradius+\highlightcurve\relax\space
           0 \strip@pt\dimexpr\highlighttotalht-\highlightradius\relax\space c
           0 \strip@pt\highlightradius\space l
           0\space \strip@pt\dimexpr\highlightradius-\highlightcurve\relax\space
           \strip@pt\dimexpr\highlightradius-\highlightcurve\relax\space 0\space
           \strip@pt\highlightradius\space 0 c
           f Q}}}%
        \raisebox{\highlightdp}{\hbox to \highlightwd{\hfil\usebox{\highlightbox}\hfil}}%
      }%
    }%
  \else
    {\setlength{\fboxsep}{1pt}\colorbox{#1}{\usebox{\highlightbox}}}%
  \fi
  \endgroup
}
\makeatother

\newenvironment{insightbox}{%
  \par\kern1.5pt
  \noindent\begingroup
  \color{gray}\vrule width 3pt%
  \hspace{10pt}%
  \color{black}%
  \begin{minipage}[t]{\dimexpr\linewidth-13pt\relax}%
}{%
  \end{minipage}%
  \endgroup
  \par\kern4pt
}

\newcommand{\myparagraph}[1]{\noindent\textbf{#1}}
\renewcommand{\email}[2][]{\ignorespaces}

\AtBeginDocument{%
  }

\copyrightyear{2026}
\acmYear{2026}
\setcopyright{cc}
\setcctype{by}
\acmConference[DaMoN '26]{22nd International Workshop on Data Management on New Hardware}{May 31-June 05, 2026}{Bengaluru, India}
\acmBooktitle{22nd International Workshop on Data Management on New Hardware (DaMoN '26), May 31-June 05, 2026, Bengaluru, India}
\acmDOI{10.1145/3789237.3809125}
\acmISBN{979-8-4007-2455-8/2026/05}

\begin{document}

\title{Do GPUs Really Need New Tabular File Formats?}

\authorOrig{Jigao Luo}
\authornoteOrig{Correspondence goes to \href{mailto:jigao.luo@tu-darmstadt.de}{\texttt{jigao.luo@tu-darmstadt.de}}.}
\orcidOrig{0009-0005-2263-1959}
\affiliationOrig{%
  \institutionOrig{TU Darmstadt}
  \cityOrig{Darmstadt}
  \countryOrig{Germany}
}

\authorOrig{Qi Chen}
\orcidOrig{0009-0003-9495-9123}
\affiliationOrig{%
  \institutionOrig{TU Darmstadt}
  \cityOrig{Darmstadt}
  \countryOrig{Germany}
}

\authorOrig{Carsten Binnig}
\orcidOrig{0000-0002-2744-7836}
\affiliationOrig{%
  \institutionOrig{TU Darmstadt \& DFKI \& hessian.AI}
  \cityOrig{Darmstadt}
  \countryOrig{Germany}
}

\renewcommand{\shortauthors}{Jigao Luo, Qi Chen, and Carsten Binnig}

\begin{abstract}

\rev{
Apache Parquet is the de facto columnar file format in modern analytical systems, yet its configuration guidelines remain largely shaped by CPU-centric execution engines. 
As GPU-accelerated data processing becomes increasingly prevalent, Parquet files optimized by default for CPU-oriented engines often severely underutilize GPU hardware, 
unnecessarily turning GPU scans into an artificial performance bottleneck.

In this work, we systematically study how Parquet configurations affect GPU scan performance, 
and show that poor Parquet performance on GPUs is mostly a consequence of suboptimal configuration choices. 
By applying GPU-aware configurations, we achieve up to 125 GB/s effective read bandwidth using fully compliant Parquet files, without switching to a new file format or sacrificing the interoperability benefits of the de facto~standard.
}

\end{abstract}

\ccsdesc[300]{Information systems~Online analytical processing engines}

\keywords{File Format, GPU, NVMe SSDs, Encodings, Compression, Parquet}


\maketitle

\section{Introduction}

\myparagraph{Parquet Dominance.}
Over the past decade, Parquet has become the de facto columnar file format in modern analytics systems. 
It is widely adopted across analytical databases and query engines 
such as DuckDB~\cite{DBLP:conf/sigmod/RaasveldtM19}, Velox~\cite{DBLP:journals/pvldb/PedreiraEBWSPHC22}, and DataFusion~\cite{DBLP:conf/sigmod/LambSHCKHS24}, 
all of which support direct querying of Parquet files.
\rev{The format exposes several} configuration \rev{parameters that affect performance, including} page counts, row group sizes, encoding, and compression. 
\rev{Yet the default parameter values} were chosen based on CPU-oriented access patterns and I/O behavior~\cite{duckdb-parquet-perf,duckdb-parquet-tips,DBLP:journals/pvldb/ZengHSPMZ23,DBLP:journals/pvldb/LiuPIH23}. 
\rev{These defaults are used for most Parquet files, even when those files will be used for GPU~scans.}

\myparagraph{GPU Bottleneck With Parquet.}
Despite extensive CPU-side testing, Parquet's \rev{reported} performance on GPUs remains subpar. 
In practice, Parquet \rev{reading} is a \rev{major} bottleneck in GPU-accelerated analytics: 
\rev{the} NVIDIA RAPIDS team reports that about \textbf{85\%(!)} of TPC-H benchmark runtime is spent on Parquet scan rather than on other query operators~\cite{veloxconf_cudf}. 
\rev{This naturally raises our central research question}: \textit{do GPUs really need new tabular file formats?}

\begin{figure}[t]
  \centering
  \includegraphics[width=\columnwidth]{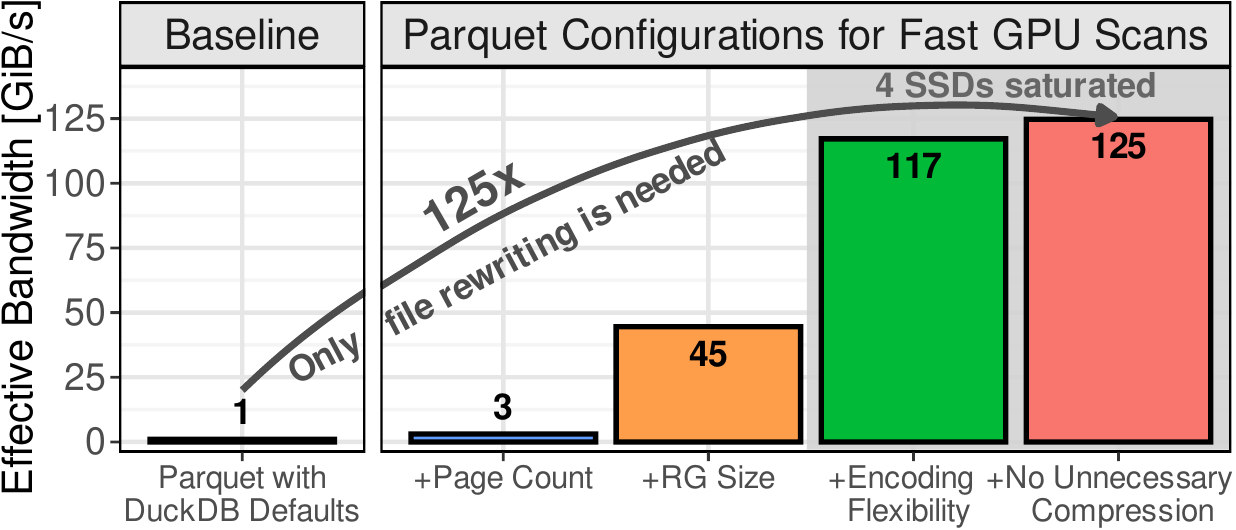}
  \caption{GPU Parquet scan on TPC-H SF300 \texttt{lineitem} with 4 SSDs: file configuration impact on effective read bandwidth.}
  \Description{Bandwidth comparison for several Parquet file configurations on a GPU scan using four SSDs.}
  \label{fig:figure1}
\end{figure}

\myparagraph{Why Is Parquet Not Optimal for GPUs?}
As shown in~\Cref{fig:figure1} (left), \rev{Parquet written with the default configuration delivers poor GPU scan performance, while GPU-aware configurations yield dramatic improvements.}
Simply by rewriting Parquet files with GPU-friendly settings, the effective bandwidth increases up to 125 GB/s. 
As such, we believe that in many cases, Parquet itself is fully capable of achieving high GPU performance without requiring a new file format \rev{for~GPUs}.

\myparagraph{Need for New File Formats?}
The goal of this work is to push the existing Parquet format to its practical limits before considering entirely new \rev{file-}format designs for GPUs. 
Such formats would need to \rev{outperform the optimized Parquet configuration} shown in~\Cref{fig:figure1}, \rev{rather than merely improving over the defaults}, a direction \rev{beyond} the scope of this work.
Our \rev{aim instead} is to thoroughly explore the performance potential \rev{already available} within the Parquet specification \rev{and to} establish a strong baseline for future GPU file formats.
Recent work has also proposed \rev{new} file formats for GPUs, but these efforts \rev{mainly target} in-GPU-memory decoding~\cite{DBLP:journals/pvldb/AfroozehB25,DBLP:conf/damon/HepkemaAFBM25,DBLP:conf/damon/AfroozehFB24,DBLP:journals/sigops/VonkHA25,futuredata-vortex}.
Their approach differs from this work in two ways. 
First, we evaluate practical scenarios by reading directly from SSDs rather than GPU memory. 
Second, we preserve the existing Parquet ecosystem while maintaining compatibility \rev{and~interoperability}.

\myparagraph{Contributions.}
In this short paper, we present a set of insights and the first systematic evaluation of how Parquet configuration choices affect GPU scan performance. 
We also provide a rewriter tool\footnote{The Parquet rewriter tool is available at: \\ \url{https://github.com/DataManagementLab/ParquetRewriter}.}
that transforms Parquet files into arbitrary configurations.
Collectively, our insights and tool provide the first \rev{practical} guidance on optimizing Parquet for GPU databases, 
enabling substantial acceleration without requiring a completely new file~format.

\section{Parquet in GPU Databases}

Apache Parquet~\cite{10.5555/3012315} is a widely used columnar \rev{file} format for analytical processing. 
A row group (RG) contains one column chunk per column, and each column chunk is divided into pages. 
Pages may use different encodings from both the older V1 and newer V2 schemes.
After encoding, compression is applied at the column chunk level using algorithms such as Snappy, Zstandard, or~\rev{GZIP}.

NVIDIA RAPIDS cuDF~\cite{libcudf} is the only GPU-accelerated library that reads Parquet directly on the GPU, with both decompression and decoding executed as \rev{GPU} kernels. 
With GPUDirect Storage (GDS)~\cite{gds_website}, cuDF reads Parquet directly from storage into GPU memory. 
As a foundational library, cuDF underpins several GPU-accelerated analytical systems, 
including Theseus~\cite{DBLP:journals/corr/abs-2508-05029}, Sirius~\cite{DBLP:conf/cidr/YogatamaYKSKCTP26}, Velox-cuDF~\cite{veloxconf_cudf}, and PystachIO~\cite{DBLP:journals/corr/abs-2512-02862}. 
Among them, PystachIO is the only system that fully overlaps I/O and GPU computation.
In this work, we use PystachIO to investigate Parquet configuration and optimizations on~GPUs.

\begin{figure}[t]
  \centering
  \includegraphics[width=\columnwidth]{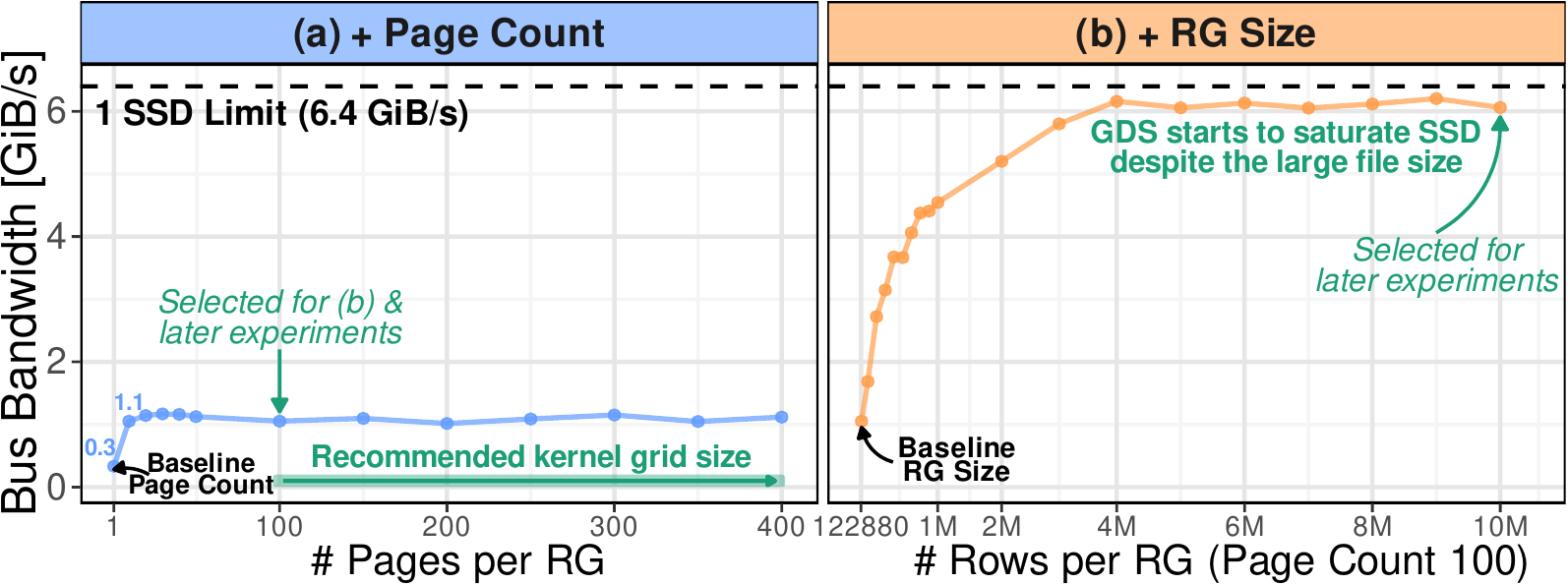}
  \caption{GPU Parquet scan on TPC-H SF300 \texttt{lineitem} with one SSD: storage bus bandwidth of different file configurations. 
  Left: varying page counts. Right: varying rows per RG.}
  \Description{Two plots showing how page count and row-group size affect GPU Parquet scan bandwidth with one SSD.}
  \label{fig:page-rg}
\end{figure}

\section{Pushing Limits of Parquet for GPUs}

\rev{In this section}, we present four key insights into Parquet configuration and its effects on GPU scans. 
Our setup uses one NVIDIA A100 GPU reading \rev{the largest} TPC-H SF300 \rev{table, \texttt{lineitem},} from local \rev{NVMe} SSDs via~GDS.
Guided by these insights, we develop a rewriter tool \rev{for simple and flexible reconfiguration of Parquet~files.}


\myparagraph{Baseline.}
The baseline Parquet file, used as the initial input to our rewriter, is generated by DuckDB using its default settings. 
Under these defaults, 
\highlight[ggbluelight]{each column chunk contains a single page}, 
\highlight[ggorangelight]{each RG contains 122880 rows}~\cite{duckdb-parquet-perf,duckdb-parquet-tips}, 
and \highlight[gggreenlight]{encodings are restricted} \highlight[gggreenlight]{to Parquet V1}.
The page count and RG size follow DuckDB's implementation practice of mapping RGs, not pages, to CPU cores for parallelism.
DuckDB defaults to older \rev{Parquet} V1 encodings to maintain broad compatibility with systems that may not support reading \rev{Parquet} V2~encodings.

\myparagraph{Increase Page Count.}
To read Parquet files on GPUs, we use cuDF, which maps the number of pages to the grid size in its GPU kernel launch parameters.
In effect, cuDF parallelizes across both RGs and pages to exploit the GPU's massive parallelism.
This leads to an important consequence: if a Parquet file contains too few pages, only a small portion of the GPU becomes active during decoding. 
To achieve high GPU utilization, the page count specified at file generation should match the kernel grid size recommended for modern GPUs -- typically well above 100.

As shown in~\Cref{fig:page-rg}(a), increasing the page count from 1 raises the read bandwidth. 
The gain comes from the decoding kernel running more efficiently. 
However, increasing the page count further does not improve storage bandwidth, a limitation we examine next.
In the following experiments, we use a page count of 100 as a reasonable choice for kernel grid size. 
There is no single universally optimal page count; the key is simply to avoid small values.

\begin{insightbox}
\noindent\highlight[ggbluelight]{\textbf{Insight 1:}}
Given the characteristics of GPU decoding kernels, increasing the page count to 100 or above is recommended.
\end{insightbox}

\myparagraph{Improve RG Size.}
The GPU I/O stack in GDS behaves quite differently from the CPU I/O stack, leading to a different choice of optimal RG size.
DuckDB determines its RG size through tailored microbenchmarking~\cite{duckdb-parquet-perf,duckdb-parquet-tips}, producing column chunks on the order of 100 KB once compressed and encoded. 
For GDS reads, however, such small I/O units are suboptimal, as MiB-scale transfers tend to be more efficient~\cite{DBLP:conf/damon/TorpLT25,DBLP:journals/pacmmod/BoeschenZB24}.
This motivates using a much larger RG size so that each column chunk reaches an I/O size that allows GDS to saturate storage.

The small I/O size implied by DuckDB's RG size explains why~\Cref{fig:page-rg}(a) stops improving; 
as shown in~\Cref{fig:page-rg}(b), increasing the RG size beyond 4M rows helps GDS saturate one SSD.
In the following experiments, we use a larger RG size of 10M rows, 
as the SSD must remain saturated even when our subsequent techniques further reduce the I/O size.
This also reinforces our \highlight[ggbluelight]{Insight 1}: 
such a large RG naturally requires fine-grained page-level parallelism.

\begin{insightbox}
\noindent\highlight[ggorangelight]{\textbf{Insight 2:}}
Given the GPU I/O stack, million-row RG sizes are preferred to allow GDS to saturate storage.
\end{insightbox}

\begin{figure}[t]
  \centering
  \includegraphics[width=\columnwidth]{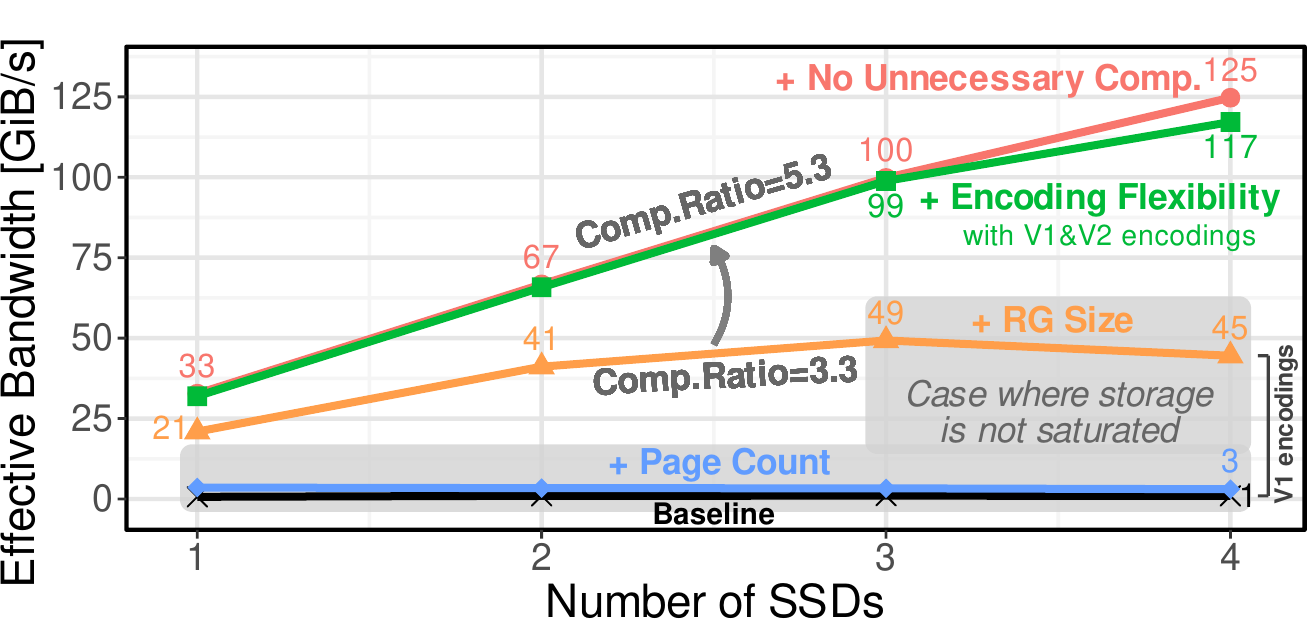}
  \caption{GPU Parquet scan on TPC-H SF300 \texttt{lineitem}: file configuration and SSD scaling effects on effective~bandwidth.}
  \Description{Ablation plot comparing effective bandwidth for Parquet configuration optimizations across SSD counts.}
  \label{fig:ablation}
\end{figure}

\myparagraph{Encoding Flexibility.}
Most Parquet writers apply a single V1 encoding to an entire column, 
unlike proprietary formats that use per-chunk \rev{encoding} strategies tailored to local data distributions~\cite{DBLP:journals/pacmmod/KuschewskiSAL23,DBLP:conf/sigmod/LangMFB0K16,DBLP:journals/pacmmod/BoeschenZB24}. 
\rev{Extending these existing ideas to our Parquet rewriter}, we allow each column chunk to explore a wider range of \rev{candidate} encodings \rev{in Parquet V1 and V2},
\rev{while finalizing} only the one that yields the smallest encoded size.
\rev{The rationale for choosing the smallest size is to alleviate the I/O bottleneck~\cite{DBLP:journals/pacmmod/BoeschenZB24,DBLP:conf/damon/NicholsonCBA25,DBLP:journals/corr/abs-2512-02862,DBLP:journals/corr/abs-2602-08190,nvidiaGTC2026,DBLP:conf/cidr/JiangNSA26}.}

\rev{
Our current encoding-selection strategy explores all Parquet valid encodings for a specific data type, 
rather than relying on the sampling-based approaches used in prior work~\cite{DBLP:journals/pacmmod/KuschewskiSAL23,DBLP:journals/pacmmod/BoeschenZB24}.
This is still practically feasible because the search space is small, with fewer than five candidate Parquet encodings for any given data type.
Moreover, the overhead of the current encoding selection is relatively low. 
We provide further discussions of this overhead and GPU encodings in~\Cref{sec:discussions}, 
while leaving sampling- and heuristic-based encoding selections to our future work.
}

As shown in~\Cref{fig:ablation}, enabling encoding flexibility improves the compression ratio, reflecting the size reduction after encoding and compression. 
Note that in~\Cref{fig:ablation}, we report \textit{effective bandwidth},
computed as the \rev{logical} raw data size after decoding and decompression divided by the scan runtime.
We switch to this metric because improving RG size alone already saturates storage bandwidth (see~\Cref{fig:page-rg}(b)).
\rev{This metric thus captures the rate at which logical raw data is scanned into GPU memory.}
With encoding flexibility, effective bandwidth scales more linearly with the number of SSDs,
\rev{especially when more than two SSDs are used, compared with the case of the RG size optimization shown in~\Cref{fig:ablation}.}

\rev{
The increase in effective bandwidth from our encoding flexibility is mainly due to two reasons. 
First, the smaller encoded size achieved by better encodings efficiently mitigates the I/O bottleneck, since less data needs to be read from storage. 
Prior Parquet V1 encodings are typically limited to plain or dictionary encoding, 
which do not fully exploit the underlying data distributions, 
whereas Parquet V2 encodings, such as delta encoding, often can.
The resulting improvement in compression ratio is also annotated in~\Cref{fig:ablation}. 
Second, Parquet V2 encodings are also efficient on GPUs in both decoding bandwidth and resource usage, thereby addressing the common criticism of Parquet encodings~\cite{columnarAI,DBLP:journals/pacmmod/KuschewskiSAL23,DBLP:journals/pvldb/ZengHSPMZ23,DBLP:conf/damon/AfroozehFB24}. 
This is also consistent with feedback from the cuDF team, 
who noted that delta encoding \texttt{DELTA\_BINARY\_PACKED} in V2 achieves high decoding bandwidth for integer data~types.
}


\begin{insightbox}
\noindent\highlight[gggreenlight]{\textbf{Insight 3:}} Flexibly choosing \rev{modern Parquet} encodings to achieve higher size reduction improves GPU scan performance.
\end{insightbox}

\myparagraph{No Unnecessary Compression.}
A practical limitation of compression arises when it is applied blindly, even in cases where it yields no size reduction.
We believe this contributes to the common criticism of Parquet \rev{compressions}~\cite{columnarAI,DBLP:journals/pacmmod/KuschewskiSAL23,DBLP:journals/pvldb/ZengHSPMZ23,DBLP:journals/pacmmod/ZengMPMPPZ25}, where decompression overhead slows down query execution. 
In practice, compression is a tradeoff between reducing I/O size in storage and adding decompression cost during reading, so it should be applied selectively rather than unconditionally.
Following this principle, we finalize compression for a column chunk only when its size reduction exceeds a chosen threshold (10\% in our experiments); otherwise, the column chunk remains uncompressed.
\rev{We choose this 10\% threshold empirically: it helps skipping cases where there is no gain, or where the gain is too small to justify the additional decompression cost.
We acknowledge that an optimal threshold should ideally be determined by considering the tradeoff between decompression bandwidth and I/O bandwidth, 
and we leave this as future~work.
}

As shown in~\Cref{fig:ablation}, avoiding unnecessary compression improves performance only when the GPU reads Parquet from four SSDs. 
This is expected: reading four SSDs, the GPU tends to become compute-bound as many decompression and decoding kernels are queued.
Removing unnecessary decompression allows \rev{other} useful kernels to run earlier. 
The effect disappears with fewer SSDs because the workload becomes I/O-bound, and the GPU has ample compute capacity for decompression.
\citeauthor{DBLP:conf/damon/NicholsonCBA25} observed similar behavior, where decompression degraded performance once storage bandwidth approached the PCIe limit~\cite{DBLP:conf/damon/NicholsonCBA25}.
Although the gains are modest, we still recommend avoiding unnecessary compression, as the file must be read efficiently across many possible~cases.


\begin{insightbox}
\noindent\highlight[ggredlight]{\textbf{Insight 4:}} Skip unnecessary compression if no size~reduction.
\end{insightbox}


\section{Overlapping I/O and GPU Computation}

\rev{
Given the optimized file configuration, a remaining question in the data path is how file-level improvements translate into gains at the query level. 
To this end, we next discuss overlapping Parquet reading with GPU query processing.
We focus on overlap because it is a key technique for mitigating I/O bottlenecks when GPUs process datasets that exceed memory capacity and are therefore transferred from storage,
and it is widely used in recent GPU systems~\cite{DBLP:journals/pacmmod/BoeschenZB24,DBLP:journals/pvldb/YuanIMT24,DBLP:journals/corr/abs-2602-08190,DBLP:journals/corr/abs-2512-02862}.

In this section, we continue to use a single NVIDIA A100 GPU with one SSD and evaluate the query runtime of our GPU query engine, PystachIO.
To this end, we study two I/O-intensive TPC-H queries: Query 6, a single-table aggregation query that essentially scans the full Parquet file, 
and Query 12, which includes a join between the two largest tables in~TPC-H.
}

\begin{figure}[t]
    \centering
    \includegraphics[width=\columnwidth]{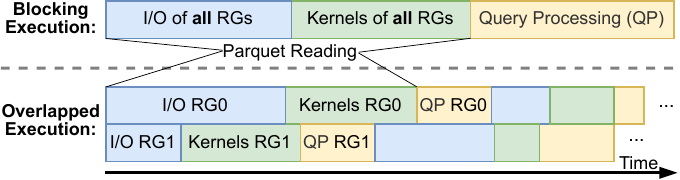}
    \caption{\rev{High-level comparison of blocking and overlapped execution. 
                  (1) The Parquet reading stage consists of storage I/O and GPU kernels. 
                  (2) Full query execution comprises Parquet reading followed by query processing~operators.}}
    \Description{Timeline diagram comparing blocking execution with overlapped Parquet reading and query processing.}
    \label{fig:query-processing}
\end{figure}

\subsection{Overlapped Parquet Reading}

\rev{
We additionally study GPU Parquet reader optimizations, with a more detailed analysis given in~\cite{DBLP:journals/corr/abs-2512-02862}. 
The key finding is that an overlapped reader is a prerequisite for fully realizing both the benefits of the optimized file configuration and the GPU performance potential.
It also serves as a necessary foundation for the performance results reported in this work.
\Cref{fig:query-processing} illustrates this through a comparison of blocking and overlapped Parquet reading. 
In the blocking design, decompression and decoding kernels can be launched only after the entire I/O phase completes, leaving the GPU entirely idle during that time.
In contrast, overlapping I/O with GPU computation not only enables kernel execution during the I/O phase but also helps avoid out-of-memory errors by processing data at RG granularity.

This work on file configuration and the study of reader design in~\cite{DBLP:journals/corr/abs-2512-02862} provide complementary perspectives, 
and both are necessary for understanding GPU Parquet scan performance. 
We therefore view this performance as determined by two tightly coupled dimensions: \textbf{file configuration} and \textbf{reader design}. 
Both are necessary, and neither alone is sufficient: even with our optimized overlapped reader, performance remains poor if the file configuration itself is unfavorable to GPUs, as illustrated in \Cref{fig:ablation,fig:scaling}.
In~\Cref{fig:scaling}, the baseline and increased page count both lead to very slow runtime, 
whereas increasing RG size brings performance much closer to the optimized file configuration, 
which incorporates the encoding and compression optimizations introduced~earlier.
}

\subsection{Overlapped Query Processing}

\begin{figure}[t]
    \centering
    \includegraphics[width=\columnwidth]{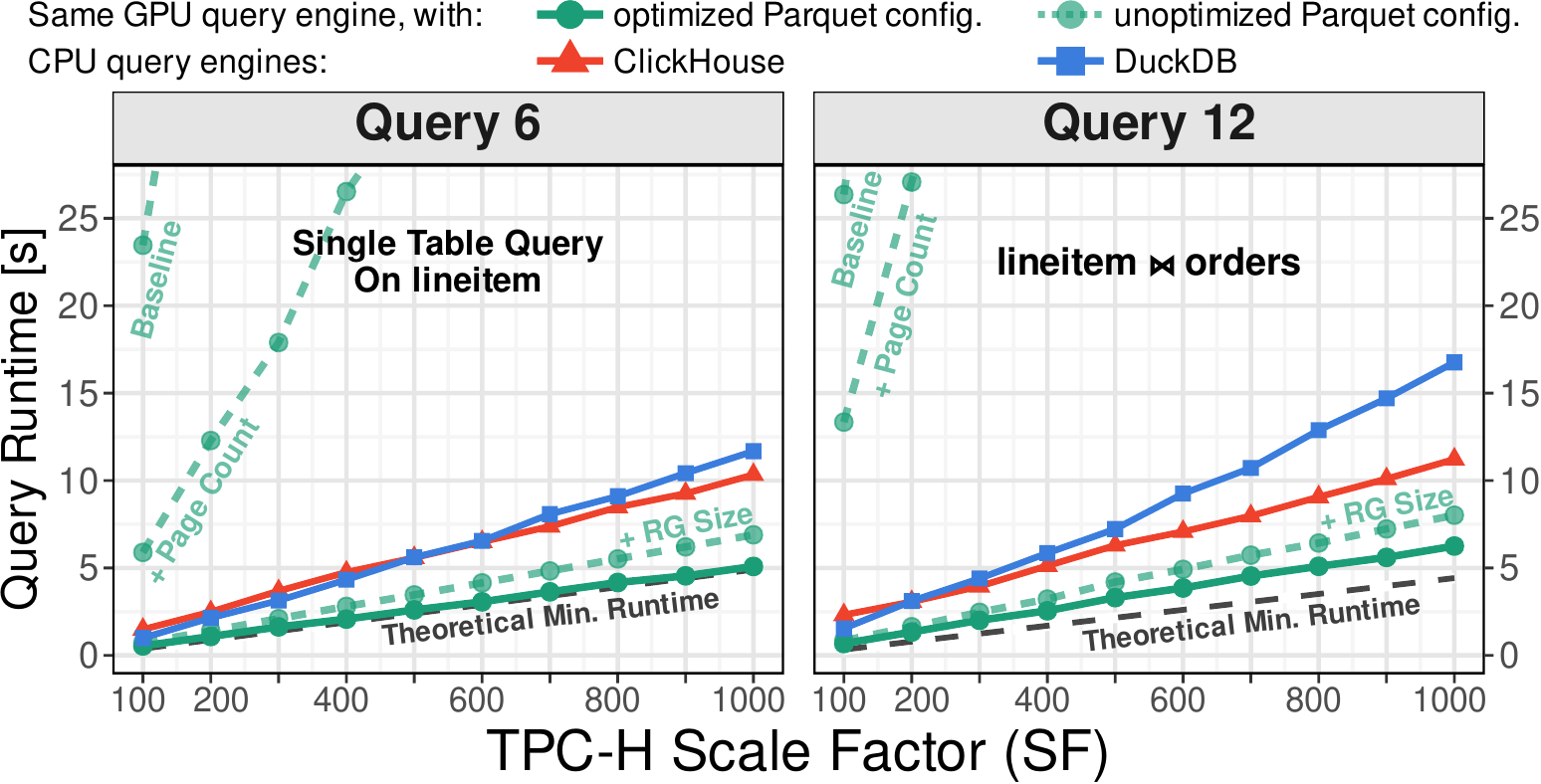}
    \caption{\rev{Dataset scalability up to TPC-H SF1000 on one SSD: query performance of the GPU query engine PystachIO. 
                 (1) Comparison across different Parquet configurations, with extremely high runtimes omitted. 
                 The optimized case combines \texttt{+Encoding Flexibility} with \texttt{+No Unnecessary Compression}. 
                 (2) Comparison against CPU systems, ClickHouse and DuckDB, on the same optimized Parquet files.}}
    \Description{Scalability plots comparing PystachIO query runtime across Parquet configurations and against CPU systems.}
    \label{fig:scaling}
\end{figure}

\rev{
With both the file configuration and the reader optimized, we further study the effect of overlap on end-to-end GPU query processing. 
A more detailed analysis is given in~\cite{DBLP:journals/corr/abs-2512-02862}. 
The key idea is to extend overlap beyond Parquet reading to additional query operators, increasing the scope of overlap between I/O and computation.
Once I/O dominates execution time, broadening the overlap helps improve query performance by further hiding I/O latency.
With overlapped query processing in~\Cref{fig:query-processing}, each RG produced by Parquet reading is directly consumed by a query operator, e.g., on the build side of a hash~join.

To demonstrate the effectiveness of overlapped GPU query processing, we compare our GPU system against two CPU systems, shown as solid lines in~\Cref{fig:scaling}. 
All three systems use the same Parquet files but different hardware setups: 
the CPU systems run on an AMD EPYC 9654P 96-Core Processor without a GPU. 
To indicate proximity to the theoretical lower bound of query runtime, 
the gray reference line is computed as the total read size from storage divided by SSD~bandwidth.
Overall, the GPU system outperforms the CPU systems by a large margin and remains close to the theoretical minimum, 
demonstrating that our file-level optimizations are effective and already translate into substantial improvements at the query level.
Since CPU system performance does not vary substantially across different file configurations,
we use the same optimized configuration for a fair comparison.
Our GPU engine remains close to the theoretical minimum for both queries.


}

\section{Discussions}
\label{sec:discussions}

\myparagraph{Overheads of Parquet Rewriting.}
By using Rust and an efficient multithreaded implementation, \rev{our} rewriter fully utilizes modern CPUs. Rewriting typically completes within minutes for a 100 GB dataset. 
\rev{
For many systems, an offline one-time preprocessing step is already common, and since rewriting often reduces file size, it introduces no additional storage~overhead.
The closest related work, the proprietary Microsoft V-Order~\cite{vorder,vorder-blog1,vorder-blog2},
reports an additional rewriting cost of 10-20\% of the original write time and targets only Parquet configurations for CPU-based~systems.
}

\myparagraph{\rev{Future File Formats for GPUs.}}
\rev{
Several GPU-friendly encodings~\cite{DBLP:journals/pvldb/AfroozehB25,DBLP:conf/damon/HepkemaAFBM25,DBLP:conf/damon/AfroozehFB24,DBLP:journals/sigops/VonkHA25} 
have been proposed to remove data dependencies to make decoding highly parallelizable on GPUs. 
These schemes report extremely high in-GPU-memory decoding throughput -- up to 5700 GB/s~\cite{columnarAI,futuredata-vortex}. 
While these results are impressive, many real-world settings require reading files first from SSDs, 
and the dominant bottleneck is still storage I/O rather than computation~\cite{DBLP:journals/pacmmod/BoeschenZB24,DBLP:conf/damon/NicholsonCBA25,nvidiaGTC2026,DBLP:journals/corr/abs-2512-02862,DBLP:journals/corr/abs-2602-08190,DBLP:conf/cidr/JiangNSA26}. 
There is also a strong interest in new file formats across both academia and industry.
Some efforts aim to replace Parquet with new file formats, mainly to support these new GPU-friendly encodings~\cite{futuredata-vortex,columnarAI,DBLP:journals/corr/abs-2504-15247}. 
At the same time, the Apache Parquet community is discussing adding these encodings~\cite{cudfIssue21173,FSSTParquet,parquetFormatPR557}. 

In this context, we present three perspectives on GPU file formats loaded from storage. 
First, for I/O-bound workloads, the extremely high in-memory decoding bandwidth offered by these encodings may provide limited benefit, 
since storage bandwidth is usually only on the order of tens of GB/s, far below GPU HBM bandwidth in the TB/s range. 
As a result, file scans are typically limited by I/O rather than by memory bandwidth or computation. 
Second, compression ratio matters in I/O-bound cases, as it directly improves effective bandwidth (see~\Cref{fig:ablation}). 
Finally, even if a new file format were introduced, 
much of the effort in this work would still be required in a similar form, 
namely, a rewriter from Parquet to the new file format. 
Doing so would also sacrifice the benefits of the Parquet ecosystem, especially compatibility and~interoperability.
}

\myparagraph{\rev{Future I/O Data Paths.}}  
\rev{
One often overlooked dimension in studying file formats is the I/O data path, namely, where the input data is stored and how it is transferred. 
In this work, we use local-attached NVMe SSDs, since this setup is the simplest and also enables us to scale the number of SSDs.
However, this is no longer the dominant configuration in the AI era, where cloud object storage~\cite{DBLP:journals/pvldb/DurnerL023} and network-attached storage~\cite{DBLP:conf/btw/LuoB0B25} accessed via NICs are becoming increasingly important for GPUs. 
This shift is also visible in NVIDIA's high-end DGX servers integrating eight GPUs, each with its own NIC, but only a single SSD~\cite{nvidiaDGXB300Guide,nvidiaDGXH100Guide}. 
We therefore expect future file-format studies to increasingly consider these storage settings.
}

\myparagraph{Rewriting Beyond GPU.} 
\rev{While we focus on GPUs, our} rewriter is not GPU specific, and applies Parquet configurations without hardware-specific assumptions. 
Rewriting Parquet for different hardware simply means selecting different configurations. 
For example, \highlight[gggreenlight]{Insight~3} and \highlight[ggredlight]{Insight~4} also apply to CPU-based systems, 
and parts of these ideas are already being discussed in the Arrow Rust library~\cite{arrow-rs-issue-7739,arrow-rs-issue-8358,arrow-rs-issue-8378,arrow-rs-pr-8257,arrowRsPR9826}. 
Overall, this leads to a simple principle: 
\rev{when a file is queried frequently, rewriting it into a hardware-optimized format can improve performance in subsequent queries. 
While this reasoning justifies new formats, we have shown it also applies to the Parquet format itself with a different~configuration.}

\begin{acks}
This work has been partially funded by the LOEWE Spitzenprofessur of the state of Hesse (III 5-519/05.00.003-(0005)), and by the Deutsche Forschungsgemeinschaft (DFG, German Research Foundation) under Germany's Excellence Strategy (EXC3057/1 “Reasonable Artificial Intelligence”, Project No. 533677015).
We also thank DFKI Darmstadt and hessian.AI for their support 
and grateful to Nils Boeschen, Andrew Lamb, and the RAPIDS team for their~feedback.
\end{acks}

\clearpage

\bibliographystyle{ACM-Reference-Format}
\bibliography{references}

\end{document}